\documentclass[11pt]{article}
\newcommand{\Comment}[1]{{}}
\usepackage{amsfonts,amsthm,amsmath,amssymb,slashed}
\usepackage[textwidth = 430 pt, textheight = 630 pt]{geometry}
\usepackage{color}

\Comment{\usepackage{color}
\definecolor{MyDarkBlue}{rgb}{0.15,0.15,0.45}
\usepackage[linktocpage=true]{hyperref}
\hypersetup{
colorlinks=true,
citecolor=MyDarkBlue,
linkcolor=MyDarkBlue,
urlcolor=MyDarkBlue,
pdfauthor={Horatiu Nastase and Jacob Sonnenschein},
pdftitle={The title},
pdfsubject={hep-th}
}

\usepackage[numbers,sort&compress]{natbib}
\usepackage{hypernat}}
\usepackage{graphicx}
\usepackage{cite}

\newcommand\ignore[1]{}
\def\one{{\,\hbox{1\kern-.8mm l}}}

\def\a{\alpha}\def\b{\beta}

\def\d{\partial}

\newcommand{\Cset}{{\,\,{{{^{_{\pmb{\mid}}}}\kern-.45em{\mathrm C}}}}}

\newcommand{\be}{\begin{equation}}
\newcommand{\bea}{\begin{eqnarray}}

\newcommand{\ee}{\end{equation}}
\newcommand{\eea}{\end{eqnarray}}

\newcommand{\non}{\nonumber \\}
\newcommand{\CR}{\non\cr}
\newcommand{\pa}{\partial}

\def\pa{\partial}

\newcommand{\aR}{{\alpha_R}}
\newcommand{\aI}{{\alpha_I}}
\newcommand{\bR}{{\beta_R}}
\newcommand{\bI}{{\beta_I}}

\begin{document}

\renewcommand{\thefootnote}{\fnsymbol{footnote}}

\makeatletter
\@addtoreset{equation}{section}
\makeatother
\renewcommand{\theequation}{\thesection.\arabic{equation}}

\rightline{}
\rightline{}




\begin{center}
{\LARGE \bf{\sc Fluid-electromagnetic helicities and knotted  solutions of the fluid-electromagnetic equations}}
\end{center} 
 \vspace{1truecm}
\thispagestyle{empty} \centerline{
{\large \bf {\sc Horatiu Nastase${}^{a}$}}\footnote{E-mail address: \Comment{\href{mailto:horatiu.nastase@unesp.br}}
{\tt horatiu.nastase@unesp.br}}
{\bf{\sc and}}
{\large \bf {\sc Jacob Sonnenschein${}^{b,c}$}}\footnote{E-mail address: \Comment{\href{mailto:cobi@tauex.tau.ac.il}}{\tt cobi@tauex.tau.ac.il}}
                                                        }

\vspace{.5cm}


\centerline{{\it ${}^a$Instituto de F\'{i}sica Te\'{o}rica, UNESP-Universidade Estadual Paulista}} 
\centerline{{\it R. Dr. Bento T. Ferraz 271, Bl. II, Sao Paulo 01140-070, SP, Brazil}}
\vspace{.3cm}
\centerline{{\it ${}^b$School of Physics and Astronomy,}}
\centerline{{\it The Raymond and Beverly Sackler Faculty of Exact Sciences, }} \centerline{{\it Tel Aviv University, Ramat Aviv 69978, Israel}}
\vspace{.3cm}
\centerline{{\it ${}^c$ Simons Center for Geometry and Physics,}} 
\centerline{{\it SUNY, Stony Brook, NY 11794, USA }} 
\vspace{1truecm}

\thispagestyle{empty}

\centerline{\sc Abstract}

\vspace{.4truecm}

\begin{center}
\begin{minipage}[c]{380pt}
{\noindent In this paper we consider an Euler fluid coupled to external electromagnetism. 
We prove that the Hopfion fluid-electromagnetic  knot, carrying fluid and electromagnetic (EM)  helicities,  
solves the  fluid dynamical
equations as well as the Abanov Wiegmann (AW) equations for helicities,
which are inspired by  the  axial-current anomaly of a Dirac fermion. 
We also find a nontrivial knot solution with truly interacting fluid and electromagnetic fields.
The key ingredients  of these  phenomena are the 
EM and fluid helicities.
An EM dual system, with a magnetically charged fluid, is  proposed and the analogs of 
the AW equations are written down. 
We  consider a fluid coupled to a  nonlinear generalizations for electromagnetism. The Hopfions are shown to be solutions of the generalized equations. 
We write down the  formalism of fluids in  
2+1 dimensions, and we  dimensionally reduce the  3+1 dimensional solutions. 
We determine the EM knotted solutions, from which we derive the fluid knots,  by applying  special  conformal transformations with imaginary parameters on un-knotted null constant EM fields.

}
\end{minipage}
\end{center}

\vspace{.5cm}

\setcounter{page}{0}
\setcounter{tocdepth}{2}

\newpage

\tableofcontents
\renewcommand{\thefootnote}{\arabic{footnote}}
\setcounter{footnote}{0}

\linespread{1.1}
\parskip 4pt



\section{Introduction}

The theory of fluids, and the theory of electromagnetism, have both a long history. An interesting set of solutions to both are 
knotted solutions, often solutions with a nonzero Hopf index, or "Hopfions" (they appear in other areas of physics as well\footnote{See 
for instance the website www.hopfion.com}), though writing explicit forms of the solutions is often challenging. 
The solutions to Maxwell's electromagnetism without sources were written by Ra\~nada in \cite{Ranada:1989wc,ranada1990knotted},
after the early work by Trautman in \cite{Trautman:1977im}. There are solutions which are null in terms of the 
Riemann-Silberstein vector $\vec{F}=\vec{E}+i\vec{B}$, i.e., $\vec{F}^2=0$, easy to describe in terms of the Bateman construction
\cite{Bateman:1915}, or even partially null solutions. These solutions are not solitonic, since electromagnetism is linear. 

On the other hand, fluid dynamics is nonlinear, so it is even more difficult to describe, so fluid knots remained an abstract, yet 
fertile ground for a long time \cite{kambe1971motion} (see the reviews \cite{ricca1996topological,ricca1998applications,ricca2009new} 
and the book \cite{arnold1999topological}). Only relatively recently we had experimental observation of fluid knots
\cite{kleckner2013creation} and numerical constructions in \cite{proment2012vortex}.
In \cite{Alves:2017ggb,Alves:2017zjt}, based on a map between electromagnetism and fluid dynamics, null pressureless fluid 
knots were obtained. 

Both electromagnetic and fluid knots also are characterized by the existence of conserved "helicities", which are spatial 
integrals of Chern-Simons like terms, in the case of electromagnetism things like 
$\int d^3x \epsilon^{ijk}A_i\d_j B_k$, obtaining helicities ${\cal H}_{ab}$, $a,b=e\ ( electric)$ or $m\ (magnetic)$
(so that ${\cal H}_{mm}=\int d^3x \vec{A}\cdot \vec{B}$, for instance), and in the case of fluid dynamics, 
${\cal H}_f=\int d^3x \epsilon^{ijk}v_i\d_j v_k=\int d^3x \vec{v}\cdot\vec{\omega}$ (thus analogous to ${\cal H}_{mm}$ in electromagnetism),
first defined by Moffat \cite{moffatt1969degree}. Some special solutions with fluid helicities were studied in 
\cite{moffatt1992helicity,moffatt1992helicity2}, and other solutions were found in \cite{ValarMorgulis,crowdy_2004,Ifidon2015216}.

In Bateman's ansatz for electromagnetism, in terms of two complex functions $\a$ and $\b$, it was shown in
\cite{Kedia:2013bw,besieris2009hopf} that one can find other solutions by replacing the pair $(\a,\b)$ with holomorphic 
transformations for them, $f(\a,\b), g(\a,\b)$. In \cite{Hoyos:2015bxa}, it was found that one can obtain 
``$(p,q)-$knotted solutions" by applying the transformations $\a\rightarrow (\a)^p,\b\rightarrow (\b)^q$ on solutions with ${\cal H}_{ee}
={\cal H}_{mm}=1$, and that such topologically nontrivial solution can be obtained from topologically trivial ones by acting with 
special conformal  transformations with {\em complex} rather than real parameters.

At the next level in complexity, one can consider fluids coupled to electromagnetism in "magnetohydrodynamics", and knots 
were considered as well, just that from the point of view of the what happens to the electromagnetic knots when they are coupled
to fluid, see for instance \cite{Kholodenko:2014apa,Kholodenko:2014wfa,Boozer,Newcomb,Irvine}.

However, an interesting case that was considered very recently by Abanov and Wiegmann \cite{Abanov:2021hio,Wiegmann:2022syo}\footnote{For earlier work considering chiral liquids, i.e., liquids with chirality with 
respect to electromagnetism, and the conservation of helicities in this context, see 
\cite{Avdoshkin:2014gpa,Kirilin:2017tdh,Mitkin:2021dxb}.}
is when an Euler fluid is coupled to {\em external} electromagnetic fields, so a case when there is no feedback from the fluid to 
electromagnetism via Ohm's law, yet the conductivity is assumed to be finite, so that the electromagnetic fields do not solve the 
vacuum Maxwell's equations. It was found that in this case, there is a {\em total} helicity that is conserved instead of individual 
fluid or electromagnetic helicities, ${\cal H}_{\rm tot}={\cal H}_f+{\cal H}_{mm}+2{\cal H}_{fm}$, and there is a chiral density and
current that is conserved, and is sourced by the anomaly $\vec{E}\cdot\vec{B}$, and that the fluid and cross helicities obey some 
equations we have dubbed Abanov-Wiegmann equations. 

In this paper, we consider the consequences of this construction for knot type solutions, and construct new knot solutions to this 
Euler fluid coupled to external electromagnetic fields. We also explore generalizations of this set-up, for instance with nonlinear
electromagnetic fields, and new helicities, as well as extending the formalism to 2+1 dimensions. We also explore the 
possibility of constructing solutions via conformal invariance. 

The paper is organized as follows. In the next section we review the AW formalism. In particular we discuss the helicities and the axial anomaly. We also present a covariant formulation.
In section 3 we show that the decoupled fluid-electromagnetic Hopfion configuration is a solution of the 
fluid and AW equations, and also find a general configuration in which fluid and electromagnetic helicities are 
truly coupled and interacting.  We start by describing knot solutions of the free Maxwell equations. 
We then provide a map from   EM knots to a fluid knots. We discuss the question of whether there are
fluid solutions with $(p,q)$ helicities. Next we prove that the fluid Hopfion configuration is a solution of the
AW equations. Finally we find the truly interacting knotted fluid-electromagnetic solution with helicities 
changing between the fluid and electromagnetic fields.
Next in section 4  we propose a magnetically charged fluid. We start with dualizing Maxwell's equations. 
We then write down the four helicities associated with  the ordinary EM equations and their duals.
Next we  derive the magnetically anologs of the AW equations and discuss possible physical systems with 
magnetically charged fluids.
Section 5 is devoted to coupling fluid to non-linear generalization of electromagnetism.  
In section 6 we discuss fluid EM systems in 2+1 dimensions and dimensionally reduce the 3+1 dimensional ones. 
In section 7 we show how to derive the EM knot solutions, which we later map to fluid knots, from constant 
null electric and magnetic fields by applying special conformal transformations with imaginary parameters. 
We conclude and  suggest several open questions in section 8.


\section{Review of 4 dimensional Euler-eletromagnetic formalism}

Consider an Euler fluid (inviscid, barotropic) composed of electrically charged particles of charge $e$ (so electrons, 
or ions), 
interacting with external Maxwell fields via the Lorentz force, as in \cite{Abanov:2021hio}. We consider that there is 
no feedback on the electromagnetic field from the fluid, so no interaction via Ohm's law $\vec{j}=\sigma \vec{E}$, though 
the conductivity $\sigma$ of the fluid is assumed to be finite (otherwise, we would have $\vec{E}+\vec{v}/c\times\vec{B}=0$, so $\vec{E}\cdot
\vec{B}=0$), i.e., 
\bea
&&\frac{d}{dt}\rho+\vec{\nabla}\cdot \rho\vec{v}=0\cr
&&(\d_t+\vec{v}\cdot\vec{\nabla})m\vec{v}+\vec{\nabla}\mu=e\vec{E}+(e/c)\vec{v}\times\vec{B}\;,\label{fluideqs}
\eea
where the first is the continuity equation, and the second is the Euler equation with a Lorentz force source,
 for an inviscid fluid, with shear viscosity $\eta=0$, and barotropic, so $p=p(\rho)$, or rather, since 
$dp=\rho d\mu$, with $\mu=\mu(\rho)$  the chemical potential, we have $\vec{\nabla}\mu=\frac{1}{\rho}\vec{\nabla}p$). Note that we will consider later also non-barotropic fluids, and then $\mu$ must be 
understood formally as $\int dp/\rho$.

\subsection{Helicities and anomaly}

The fluid helicity is normally defined as 
\be\label{Helicityf}
{\cal H}_f=\frac{1}{\Gamma^2}\int d^3x \vec{v}\cdot\vec{\nabla}\times \vec{v}=\frac{m^2}{h^2}\int d^3x \vec{v}\cdot \vec{\omega}\;,
\ee
where $\Gamma$ is a normalization constant, here taken to be $=h/m$ in order for ${\cal H}$ to be integer valued 
(which is needed for the case of the superfluid), and $\omega=\vec{\nabla}\times \vec{v}$ is the vorticity.\footnote{Note that  
\cite{Abanov:2021hio} define it with an $m$, so 
$\omega=m\vec{\nabla}\times \vec{v}$, but we use the usual definition} This helicity is conserved, $\frac{d}{dt}{\cal H}_f=0$, for a fluid without 
external sources. 

But in the case of a fluid with electromagnetic sources, as above, \cite{Abanov:2021hio} argue that we should replace the 
momentum $\vec{p}=m\vec{v}$ inside ${\cal H}_f$ with the canonical momentum 
\be
\vec{\pi}=m\vec{v}+\vec{A}\;,
\ee
to obtain the {\em total} (fluid plus electromagnetic, specifically magnetic-magnetic) helicity,
\bea\label{helicitytot}
{\cal H}_{\rm tot}&=&\frac{1}{h^2}\int d^3 x\vec{\pi}\cdot \vec{\nabla}\times \vec{\pi}=\frac{1}{h^2}\int d^3x \left[m^2\vec{v}\cdot \vec{\omega}
+\vec{A}\cdot \vec{B}+2m\vec{v}\cdot \vec{B}\right]\cr
&=&{\cal H}_f+{\cal H}_{mm}+2{\cal H}_{fm}\;,
\eea
where $\int d^3x \vec{v}\cdot\vec{B}=\int d^3x \vec{A}\cdot\vec{\omega}$ (by partial integration), so the 2 cross-terms 
("cross-helicities") are equal, giving what we called ${\cal H}_{fm}$. Note that 
\be
{\cal H}_{mm}=\int d^3x \vec{A}\cdot\vec{B}=
\int d^3x \epsilon^{ijk}A_i\d_k A_k\label{Hmm}
\ee
is the electromagnetic helicity of magnetic-magnetic type.  This total helicity is found to be conserved, 
\be
\frac{d}{dt}{\cal H}_{\rm tot}=0.
\ee

From now on, we put $h=c=1$.

The proof is easiest in a formulation in terms of 4-dimensional objects (though not Lorentz invariant) to be studied next. 
However, Abanov and Wiegmann derive the following equations for the  fluid- and cross-helicity densities, what we 
will call Abanov-Wiegmann equations in the following:
\bea
&&\d_t(m^2\vec{v}\cdot\vec{\omega})+\vec{\nabla}\cdot\left[\vec{v}(m^2\vec{v}\cdot\vec{\omega})
+m\vec{\omega}\left(\mu-\frac{m\vec{v}^2}{2}\right)+m\vec{v}\times(\vec{E}+\vec{v}\times \vec{B})\right]\cr
&&-2m\vec{\omega}(\vec{E}
+\vec{v}\times \vec{B})=0\cr
&&\d_t(m\vec{v}\cdot\vec{B})+\vec{\nabla}\cdot \left[\vec{v}(m\vec{v}\cdot\vec{B})+\vec{B}\left(\mu-\frac{m\vec{v}^2}{2}\right)
-m\vec{v}\times (\vec{E}+\vec{v}\times \vec{B})\right]\cr
&&+m\vec{\omega}(\vec{E}+\vec{v}\times \vec{B})=\vec{E}\cdot\vec{B}\;,\cr
&&\label{helicityeqs}
\eea
from which one finds the "local anomaly equation", 
\be
\dot\rho_A+\vec{\nabla}\cdot\vec{j}_A=2\vec{E}\cdot\vec{B}\;,
\ee
where the "fluid chirality density" is defined as
\be
\rho_A=m\vec{v}\cdot(m\vec{\omega}+2\vec{B})=\frac{{\cal H}^d_{\rm tot}-{\cal H}^d_{mm}}{m}\;,
\ee
i.e., the difference between the total helicity density and the magnetic-magnetic helicity density, while 
the fluid chirality current $\vec{j}_A$ is found to be
\be
\vec{j}_A=\rho_A\vec{v}+(m\vec{\omega}+2\vec{B})\left(\mu-\frac{m\vec{v}^2}{2}\right)+m\vec{v}(\vec{E}+\vec{v}\times\vec{B}).\label{vectcurr}
\ee

\subsection{4-dimensional formalism}

The equations are much easier to derive in a 4 dimensional formulation, though without Lorentz invariance. 

One first can check that the Euler equation, the second equation in (\ref{fluideqs}), can be rewritten as 
\be
\rho(\dot{\vec{\pi}}-\vec{\nabla}\pi_0)-\rho\vec{v}\times (\vec{\nabla}\times \vec{\pi})=0\;,
\ee
where $\pi_0$ is the Bernoulli function, 
\be
\pi_0=\Phi+A_0\;,\;\;\;-\Phi=\mu+\frac{m\vec{v}^2}{2}.
\ee

Then, defining the 4-current $j^\mu=(\rho,\rho v^i)$ and canonical 4-momentum $\pi_\mu=(\pi_0,\pi_i)$, we see that the above rewriting 
of the Euler equation, can be compactly rewritten as
\be
j^\mu \Omega_{\mu\nu}=0\;,\;\; \Omega_{\mu\nu}\equiv \d_\mu \pi_\nu-\d_\nu\pi_\mu\;,
\ee
which is seen to be understood in form language. Indeed, the $i$ component of this is the rewritten Euler equation, and the 0 
component is $v^i$ times the same (taking into account that the second term vanishes, 
since we have $\epsilon^{ijk}v_iv_j$). 

Then the helicity density 3-form is 
\be
h=\pi \wedge d\pi=\pi \wedge \Omega.
\ee

Its components are the total helicity density, 
\be
{\cal H}_{\rm tot}^d=h_0=\vec{\pi}\cdot (\vec{\nabla}\times \vec{\pi})
=\rho_A+\vec{A}\cdot\vec{B}\;,
\ee
and the total helicity flux,
\be
\vec{h}=\vec{\pi}\times(\dot{\vec{\pi}}-\vec{\nabla}\pi_0)-\pi_0(\vec{\nabla}\times \vec{\pi})=h_0\vec{v}-(\vec{\nabla}\times\vec{\pi})
(\vec{\pi}\cdot\vec{v}+\pi_0).
\ee

Then the conservation of the total helicity density is almost trivial in form language, 
\be
h=\pi\wedge d\pi=\pi\wedge d\Omega\Rightarrow dh=\Omega\wedge \Omega=0\;,
\ee
or in components, the continuity equation, 
\be
\dot h_0+\vec{\nabla}\cdot\vec{h}=0.
\ee

Integrating it over space with vanishing boundary conditions at infinity and using Gauss's law, we find the conservation law 
$\frac{d}{dt}{\cal H}_{\rm tot}=0$.

To find the chirality equation, consider the fluid chirality current form (based on the extension of the 0th components, the densities)
\be
j_A=\pi \wedge d\pi-A\wedge dA=(\pi -A)\wedge (d\pi+dA).
\ee

Then  this gives the 3-vector fluid chirality current,
\bea
j_A^i&=& \epsilon^{i0jk}\left[(\pi_0-A_0)[(d\pi)_{jk}+(dA)_{jk}]-(\pi_j-A_j)[(d\pi)_{0k}+(dA)_{0k}]\right]\cr
&=&-\left[-\epsilon^{ijk} mv^j(m\dot v ^k-\nabla^k\Phi-2E^k)+\Phi \epsilon^{ijk}(m\d_j v_k+2(dA)_{jk})\right]\;,
\eea
so
\be
\vec{j}_A=m\vec{v}\times(m\dot{\vec{v}}-\vec{\nabla}\Phi-2\vec{E})-\Phi(m\vec{\omega}+2\vec{B}).
\ee

Then, by substituting $m\dot{\vec{v}}$ from the Euler equation, we indeed obtain the 3-vector fluid chirality current (\ref{vectcurr}).

\section{Hopfion knotted solutions of the fluid + electromagnetism and Abanov-Wiegmann  equations}

We want to find solutions to the Euler+eletromagnetism equations, as well as to 
the  Abanov-Wiegmann equations for the helicities. In particular, we would like to find knotted solutions, 
that have a nonzero Hopf number, which is usually associated with a nonzero 
electromagnetic magnetic-magnetic helicity
(see for instance \cite{Hoyos:2015bxa,Arrayas:2017sfq}). 

\subsection{Electromagnetic knots}

In \cite{Alves:2017ggb,Alves:2017zjt} a map was given between electromagnetism and a null ($\vec{v}^2=1$) pressureless 
fluid, which was used to map the electromagnetic knot into a fluid knot. 

The electromagnetic knot in the Bateman formulation is written in terms of the complex 
Riemann-Silberstein vector 
\be
\vec{F}=\vec{E}+i\vec{B}\;,
\ee
with the Bateman ansatz
\be
\vec{F}=\vec{\nabla}\a\times\vec{\nabla}\b\;,
\ee
with the two complex scalar fields $\a,\b\in\mathbb{C}$. 

In components, the ansatz is 
\bea
E^i&=& \epsilon^{ijk} \left(\pa_j\aR\pa_k\bR- \pa_j\aI\pa_k\bI\right)\cr
B^i&=& \epsilon^{ikj} \left(\pa_j\aR\pa_k\bI- \pa_j\aI\pa_k\bR\right)\;,\label{compEB}
\eea
where the indices $I$ and $R$ refer to the imaginary and real parts, respectively.

The vacuum Maxwell's equations of motion in terms of $\vec{F}$ are 
\be
\vec{\nabla}\cdot\vec{F}=0\;;\;
\d_t\vec{F}+i\vec{\nabla}\times \vec{F}=0\;,\label{MxF}
\ee
where the first one is  trivially satisfied by the Bateman ansatz and the second takes the form 

\be
i\nabla\times (\pa_t\alpha\nabla\beta-\pa_t\beta\nabla\alpha)=\nabla\times \vec F\;,
\ee
which is satisfied if
\be
(\pa_t\alpha\nabla\beta-\pa_t\beta\nabla\alpha)=\vec F.
\ee

Solutions to this equation have necessarily a zero norm
\be\label{null}
F^2=(\pa_t\alpha\nabla\beta-\pa_t\beta\nabla\alpha)(\vec{\nabla}\a\times\vec{\nabla}\b)=0 \;,
\ee
which implies that 
\be\label{nullEB}
\vec E\cdot \vec B=0 \qquad E^2-B^2=0.
\ee

Topological  non-trivial solutions of  Maxwell's equations  that 
are characterized by non-trivial helicity  ${\cal H}_{mm}$ defined in (\ref{Hmm})  were found in \cite{Ranada:1989wc,ranada1990knotted}. 
The basic solution which carries ${\cal H}_{mm} =1$ is given by 
\be
\a=\frac{A-iz}{A+it}\;,\;\;
\b=\frac{x-iy}{A+it}\;,\;\;
A=\frac{1}{2}(x^2+y^2+z^2-t^2+1).\label{absol}
\ee

It is easy to check that if $\alpha(x^\mu), \beta(x^\mu)$ is a solution of Maxwell equation then also 
\be
g(\alpha(x^\mu),\beta(x^\mu)), h(\alpha(x^\mu),\beta(x^\mu))
\ee
for $g,h$ holomorphic functions are solutions.
In particular if a solution with $\alpha(x^\mu), \beta(x^\mu)$ carries a (1,1) helicity charges  
$({\cal H}_{mm}, {\cal H}_{ee})$ defined in (\ref{Hmm}),(\ref{Hee}),
\be
g(\alpha)=\alpha(x^\mu)^m , h(\beta(x^\mu))=(\beta(x^\mu)^n.
\ee
 are knotted solutions with charges $(m,n)$.

One can write the knotted solutions also in terms of the electromagnetic 2-form field strength and its dual as (see, e.g. \cite{Arrayas:2017sfq})
\bea
F&=&\frac{1}{2}F_{\mu\nu}dx^\mu\wedge dx^\nu=\frac{1}{4\pi i}\frac{\d_\mu \bar\phi\d_\nu \phi-\d_\nu\bar\phi\d_\mu \phi}{(1+|\phi|^2)^2}
dx^\mu\wedge dx^\nu\cr
\tilde F&=&\frac{1}{2}\tilde F_{\mu\nu}dx^\mu\wedge dx^\nu=\frac{1}{4\pi i}\frac{\d_\mu \bar\theta\d_\nu \theta-\d_\nu\bar\theta\d_\mu 
\theta}{(1+|\theta|^2)^2}dx^\mu\wedge dx^\nu\cr
\phi&=&\frac{Az+t(A-1)+i(tx-Ay)}{Ax+ty+i(A(A-1)-tz)}\cr
\theta&=&\frac{Ax+ty+i(Az+t(A-1))}{tz-Ay+i(A(A-1)-tz)}\;.\label{solF}
\eea

\subsection{ From EM knots to fluid knots}
In \cite{Alves:2017ggb,Alves:2017zjt}  a derivation of fluid knots was worked by implementing  a map from the EM knot  solutions.
This map is in fact a map between the energy momentum tensor of the EM theory  $T^{(EM)}_{\mu\nu}$ and that of   a perfect  fluid $T^{(fluid)}_{\mu\nu}$. The latter takes the well known form
\be
T^{(fluid)}_{\mu\nu}= \rho(u_\mu u_\nu) + P( g_{\mu\nu} + u_\mu u_\nu)\;,
\ee
where $\rho$ is the fluid density, $u^\mu$ is the four velocity vector and $g_{\mu\nu}$ is the space-time metric.

The conservation of the energy momentum tensor takes the form
\be
\pa^\mu T_{\mu\nu} =0\;,
\ee
which implies that 
\be
u_\mu u_\nu \pa^\mu( \rho + P) + ( \rho + P)(\pa^\mu u_\mu) u_\nu + ( \rho + P)(\pa^\mu u_\nu) u_\mu + \pa_\nu P =0 .
\ee

It turns out, as will be shown below, that the velocity four vectors that we will get from the EM 
knot configurations are null, namely $u^\mu\rightarrow v^\mu$,
\be
v^\mu v_\mu =0; \qquad v^\mu = (1, \vec v); \qquad (\vec v)^2 =1.
\ee

For such velocity four-vector the conservation of motion of $T_{\mu\nu}$ reduce to the 
continuity and Euler equations:
\be
\pa_t \rho + \vec \nabla\cdot ( \rho \vec v)=0\;,
\ee
\be
\rho \pa_t \vec v + \rho ( \vec v\cdot \vec \nabla)\vec v + \vec\nabla P =0.
\ee

For fluid with  $\vec\nabla P =0$,  one gets exactly the same equations provided that we take the following map:
\bea\label{mapEMf}
T_{00}&=& \frac{1}{2}(\vec{E}^2+\vec{B}^2) \leftrightarrow \qquad \rho\cr
T_{0i}&=& [\vec{E}\times \vec{B}]_i \leftrightarrow \qquad \rho v_i
\eea
\be
T_{ij}= - \left[ E_i E_j+ B_i B_j -\delta_{ij}\frac12( \vec E^2 + \vec B^2) \leftrightarrow \qquad  \rho v_i v_j\right].
\ee 

Solving this for $\vec{v},\rho$, we obtain
\be
\rho\leftrightarrow \frac{1}{2}(\vec{E}^2+\vec{B}^2);\;\;\;\;
v_i\leftrightarrow\frac{[\vec{E}\times \vec{B}]_i}{\frac{1}{2}(\vec{E}^2+\vec{B}^2)}\;,\label{velmap}
\ee
which is valid only under the null condition $\vec{E}\cdot\vec{B}=0$, $\vec{E}^2-\vec{B}^2=0$, satisfied 
by the electromagnetic knot from the previous subsection.

It is now easy to check that for null EM fields that obey
 $\vec{E}\cdot\vec{B}=0$, $\vec{E}^2-\vec{B}^2=0$, we have
$\vec v^2 =1 $, as follows:
\be
v_iv ^i=\frac{[E\times B]_i [E\times B]^i}{\rho^2}= \frac{ \epsilon_{ijk}E^jB^k
\epsilon^{ilm} E_lB_m}{E^2 B^2}= \frac{E^2B^2-(\vec E\cdot \vec B)^2}{E^2 B^2 }= 1.
\ee

We recapitulate that $\rho$ and $\vec v$ that follow from this map from the null 
EM fields obey the continuity and usual Euler equations.
 
In terms of the Riemman-Silberstein vector $\vec F$,  the velocity vector is given by
\be
\vec v =\frac{{\rm Im}( (\vec F)^* \times \vec F)}{|F^2|}=\frac{1}{i}\frac{ (\vec F)^* \times \vec F}{|F^2|} =\frac{[\vec{E}\times \vec{B}]_i}{\frac{1}{2}(\vec{E}^2+\vec{B}^2)}\;.
\ee 

We can now express also the vorticity in terms of $\vec F,\vec F^*$ as follows:
\be
w^i = \epsilon^{ijk} \pa_j v_k= \frac{\pa_j F{^i}F^{*j}-\pa_j F^{*i}F^j }{ |F|^2}
- \frac{(F^{*i}F^j-F^{*j}F^i)}{ (|F|^2)^2}.
\ee

The total helicity of the system  (\ref{helicitytot}) is built from three terms
\be
{\cal H}_{total}= {\cal H}_{mm}+ {\cal H}_{fm}+ {\cal H}_f
\ee

By construction the fluid that follows from the EM knot has a non-trivial ${\cal H}_{mm}$. 
On the other hand, since $\vec v\cdot \vec B=0$, we have
${\cal H}_{fm}=0$.

The fluid helicity ${\cal H}_f$  was defined in (\ref{Helicityf})  and was argued to be conserved under 
certain conditions (while the total helicity is always conserved).  
The conservation follows from Euler's equation; for  a fluid without pressure and external force,
\be
\pa_t ( \vec v\cdot \vec w) = \vec \nabla\cdot \left [\frac12 ( \vec v)^2  \vec w 
- \vec v( \vec w\cdot \vec v) \right ].
\ee

This implies that indeed ${\cal H}_f$ is conserved, 
provided the surface term in the corresponding integral vanishes. 
To check the conservation of the fluid helicity, we can substitute the expressions for 
$\vec v$ and $\vec w$ in terms of $\vec F,\vec F^*$, obtaining
\be
\vec v\cdot \vec w= \frac{ (\vec F)^* \times \vec F}{|F^2|}\vec \nabla\times \left[\frac{ (\vec F)^* 
\times \vec F}{|F^2|} \right ]=\frac{\epsilon_{ilm} F^{*l}F^m(\pa_j F^{*i}F^j-\pa_j F^{i}F^{*j})}{|F^2|}.
\ee

Next we  integrate this result, and apply $\pa_t$, and we find that  it yields a surface term.

\subsection{The fluid Hopfion}

Under the  map (\ref{mapEMf}), the basic EM Hopfion solution (\ref{absol})
 transforms into a  fluid knot solution, taking the  form 
\bea
&&v_x=\frac{2(y+x(t-z))}{1+x^2+y^2+(t-z)^2}\;,\;\;
v_y=\frac{-2(x-y(t-z))}{1+x^2+y^2+(t-z)^2}\;,\cr
&&v_z=\pm \sqrt{1-v_x^2-v_y^2}=\pm \frac{1-x^2-y^2+(t-z)^2}{1+x^2+y^2+(t-z)^2}\;,\cr
&&\rho=\frac{16(1+x^2+y^2+(t-z)^2)^2}{\left(t^4-2t^2(x^2+y^2+z^2-1)+(1+x^2+y^2+z^2)^2\right)^3}\;.\;\;\; \label{rhov}
\eea

\begin{figure}[t!]
\begin{tabular} {c}
 \includegraphics[width=7cm]{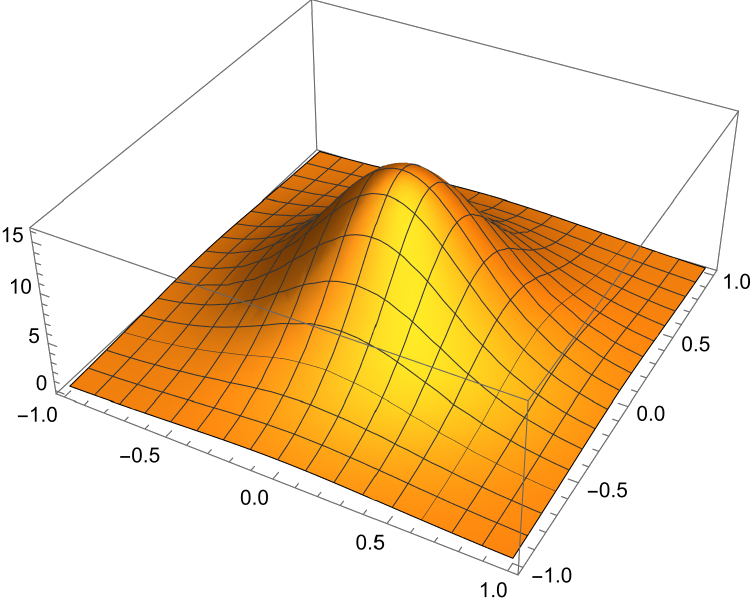}
\end{tabular}
\caption{The density of the basic (1,1)  fluid  Hopfion as a function of $x$ and $y$ for $t=0, z=0$.}  
\label{figrho11xy}
\end{figure}

The velocity profiles at $t=0$ for $z=0$ and $x=0$ are drawn in Fig.\ref {fighopf},
 and the the density at $t=0$ for $z=0$ in Fig.\ref {figrho11xy}.

\begin{figure}[t!]
\begin{tabular} {c}
\includegraphics[width=7cm]{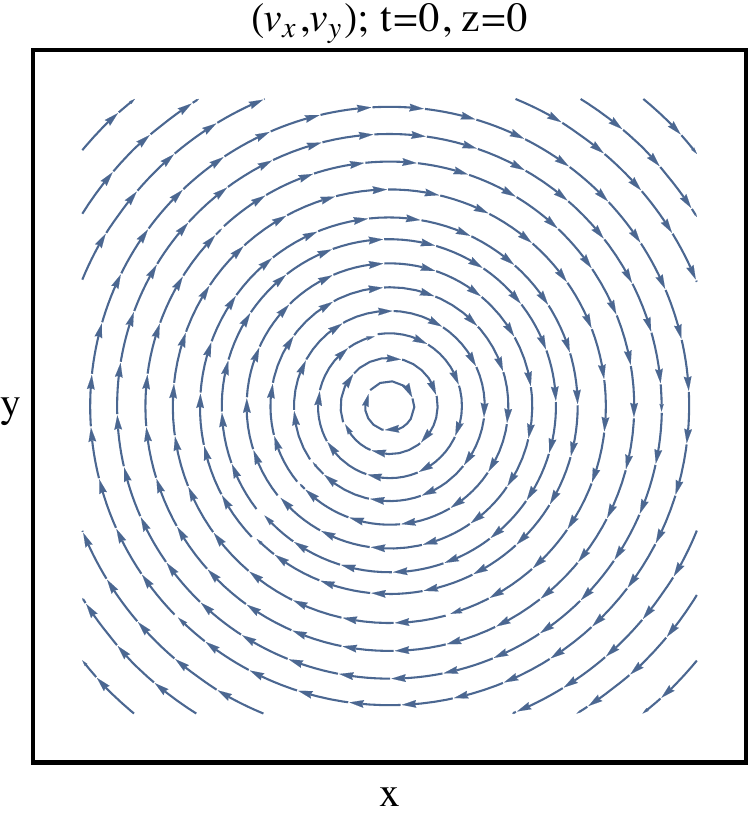}  \includegraphics[width=7cm]{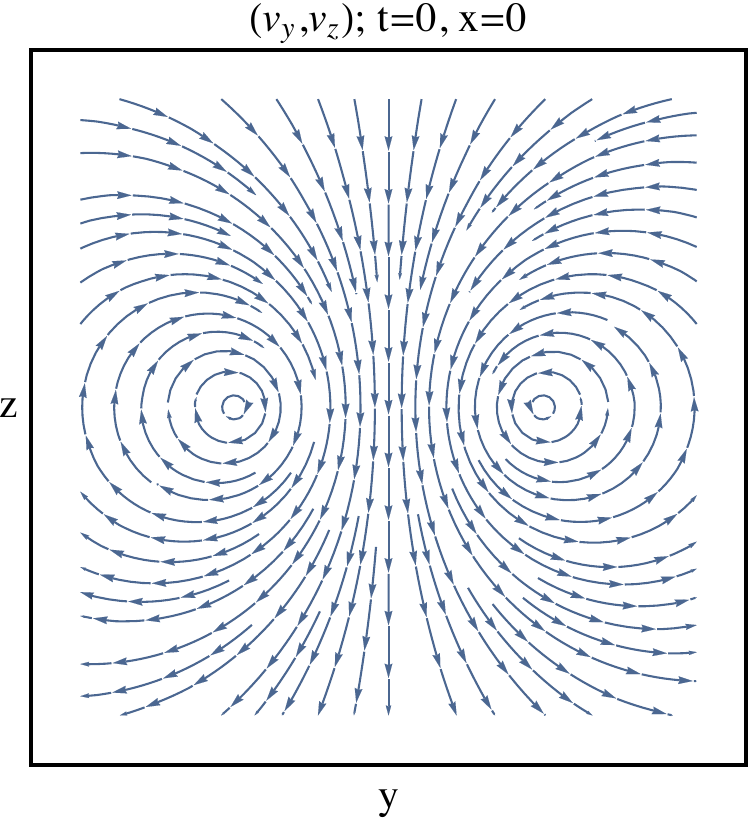}
\end{tabular}
\caption{Orthogonal sections of the velocity field for the Hopfion solution, on the $(x,y)$ plane (top) and $(y,z)$ plane (bottom).  
Using rotational symmetry in the $(x,y)$ directions the linked torus structure is apparent \cite{Alves:2017ggb,Alves:2017zjt}.}
\label{fighopf}
\end{figure}

Next we can determine the vorticity vector at $t=0$,
\be
w_x= \frac{2 \left(y \left(x^2+z^2+3\right)-2 x
   z+y^3\right)}{\left(x^2+y^2+z^2+1\right)^2}\;,
\ee
\be
w_y=\frac{-2 \left(x \left(y^2+z^2+3\right)-2 y
   z+x^3\right)}{\left(x^2+y^2+z^2+1\right)^2}\;,
\ee
\be
w_z=-\frac{4 \left(z^2+1\right)}{\left(x^2+y^2+z^2+1\right)^2}.
\ee

The vorticity has a norm of 
\be
(\vec w)^2= \frac{4 \left(x^2+y^2+4\right)}{\left(x^2+y^2+z^2+1\right)^2}\;,
\ee
and the density of the fluid  helicity is
\be
\rho_w=\vec v\cdot \vec w=\frac{4}{x^2+y^2+z^2+1}.
\ee

Next we calculate explicitly the vorticity $\omega$ and $h_0^f$ at nonzero times. 
We find
\bea
w_x &=& \frac{2 \left(t^2 y+2 t (x-y z)+y \left(x^2+z^2+3\right)-2 x z+y^3\right)}{\left((t-z)^2+x^2+y^2+1\right)^2}\CR
w_y &=& -\frac{2 \left(x \left((t-z)^2+y^2+3\right)+2 y (z-t)+x^3\right)}{\left((t-z)^2+x^2+y^2+1\right)^2}\CR
w_z &=& -\frac{4 \left((t-z)^2+1\right)}{\left((t-z)^2+x^2+y^2+1\right)^2}.
\eea

 The helicity density at nonzero times comes out to be 
\be
\rho_w=\vec{v}\cdot{w}= \frac{4}{(t-z)^2+x^2+y^2+1}.
\ee

It turns out that the fluid helicity, which is the  space integral of this density,
\be
{\cal H}= \int d^3 x\; \vec v\cdot \vec w\;,
\ee
 diverges.

If one ``normalize'' the result by subtracting  from the velocity components their  
asymptotic values, namely
\be
\vec v_n= \vec v-\vec v_{asym}\;,
\ee
one gets that the helicity density ( at t=0) takes the form 
\be
-\frac{4 \left(x^2 \left(z^2-1\right)+y^2
   \left(z^2-1\right)+\left(z^2+1\right)^2\right)}{\left(x^2+y^2+z^2+1\right)^2
   \sqrt{\left(x^2+y^2+z^2-1\right)^2+4 z^2}}
\ee
The corresponding space integral, the helicity, is still divergent.

It is interesting to note that 
 if we define a deformed  helicity density of the form  $\vec p\cdot \vec w$, 
then the correspondence helicity comes out to be finite
\be
{\cal H}_d= \int d^3 x \;\vec p\cdot \vec w= \frac{72 \pi^2}{5}.
\ee

This result is based on the values of $\vec p$ and $\vec w$ at $t=0$,
but as expected it is not conserved in time.

\subsection{Are there fluid knot solutions with higher $(p,q)$?}

Next, in analogy to the higher $(p,q)$ knots of the EM solution we would like to explore the 
possibility of having also higher fluid knot solutions  with higher $(p,q)$ knot numbers. 
For concreteness  we analyze  the (2,3) case.  We start with the Bateman configuration 
that corresponds to the EM (2,3) knot, namely
\be
\alpha=\frac{\left(-t^2+x^2+y^2+(z+i)^2\right)^2}{\left(-t (t-2
   i)+x^2+y^2+z^2+1\right)^2}
\ee
\be
\beta= \frac{8 (x-i y)^3}{\left(-t (t-2 i)+x^2+y^2+z^2+1\right)^3}.
\ee

Using (\ref{compEB}) we determine  the electric and magnetic fields, which  are written down in appendix A.
It is straightforward to check explicitely  that indeed $\vec E\cdot\vec B=0$ and $\vec{E}^2= \vec{B}^2$.

Next following the steps of (\ref{velmap}) we determine the density and the  velocity vector.
At $t=0$ the density  takes the form
\be
\rho(\vec x)= E^2=B^2 = \frac{9216 \left(x^2+y^2\right)^2 \left(2 z^2
   \left(x^2+y^2+1\right)+\left(x^2+y^2-1\right)^2+z^4\right)}{\left(x^2+y^2+z^2+
   1\right)^{10}}
\ee 

The density as a function of $(x,y)$ for $z=0$ and of $(x,z)$ for $y=0$ 
are given in Figs.\ref{fighopf} a and b, respectively.
\begin{figure}[t!]
\begin{tabular} {c}
\includegraphics[width=7cm]{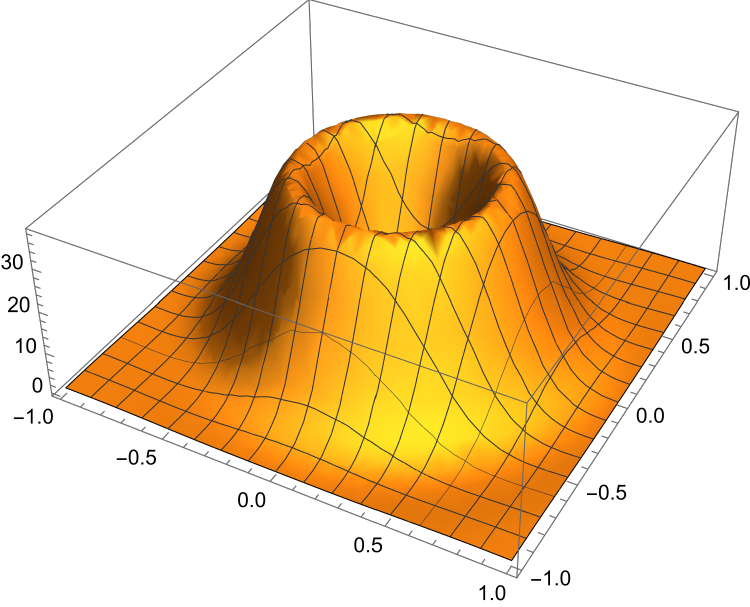}  \includegraphics[width=7cm]{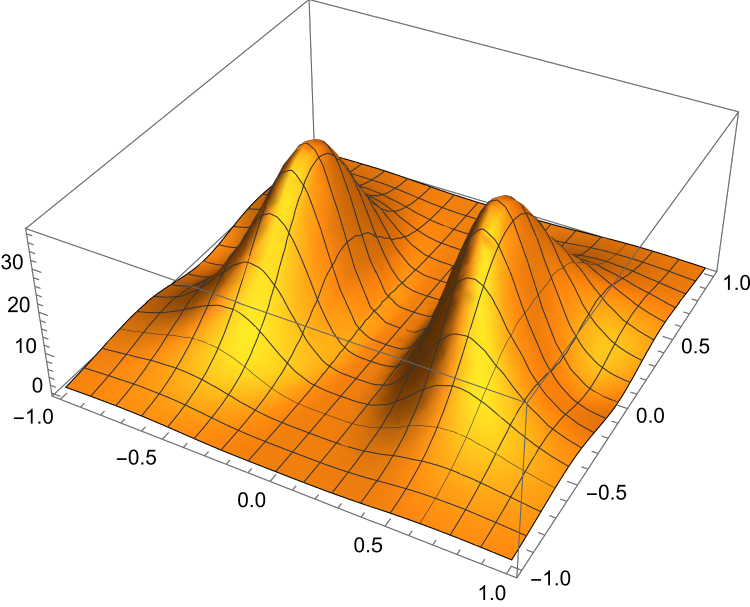}
\end{tabular}
\caption{The density of the (2,3) fluid  as a function of $(x,y)$ for $z=0$ left and of $(x,z)$ for $y=0$ on the right.} 
\label{fighopf}
\end{figure}
 
Here we faced  a surprise. Whereas the density of the (2,3) solution is different from the basic solution, 
the velocity components are identical to those of the (1,1) basic fluid knot. In fact  it is easy to show that the velocity vectors of 
all the $(p,q)$ knots are the same. This  follows  from the definition of the velocity vector (\ref{velmap})
under the substitution of $\alpha^p$, $\beta^q$.
\be
\vec v_{(pq)}=\frac{1}{i}\frac{ (\vec F)^* \times \vec F}{|F^2|}
=\frac{(\nabla \alpha^p\times \nabla\beta^q)^*\times (\nabla \alpha^p\times \nabla\beta^q)}{(\nabla 
\alpha^p\times \nabla\beta^q)^*\cdot (\nabla \alpha^p\times\nabla\beta^q)} =\frac{(\nabla \alpha\times 
\nabla\beta)^*\times (\nabla \alpha\times \nabla\beta)}{(\nabla \alpha\times \nabla\beta)^*\cdot (\nabla 
\alpha\times\nabla\beta)}=\vec v_{(11)}.
\ee

Thus the $(p,q)$ knot is characterized by its density but not by the helicity, which is the same for all of 
them.


\subsection{Fluid+Electromagnetic Hopfion as a solution to electromagnetic coupled fluid equations}

We have seen that the fluid Hopfion (\ref{rhov}) is a solution of the usual Euler equations and 
the continuity equation, and the electromagnetic Hopfion is a solution of the Maxwell's equations without source. 
But we want a solution of the continuity and Euler equation for the electromagnetic coupled fluid (\ref{fluideqs}), 
with nonzero $(\vec{v},\vec{E},\vec{B})$. 

As a first possibility, we consider the case that the fluid $\vec{v}$ follows via the map (\ref{mapEMf}) 
from the same $\vec{E},\vec{B}$ that 
appears on the right-hand side (as a source) of the fluid equations (\ref{fluideqs}). The map leads to a 
null velocity field, $v_\mu v^\mu=0$, or $v_iv^i=1$.

But under the null condition, for the fluid $\vec{v}$ interacting with electromagnetism $\vec{E},\vec{B}$, 
we get 
\be
v^i=\frac{\epsilon^{ijk}E_j B_k}{\vec{B}^2}\Rightarrow
-(\vec{v}\times \vec{B})^i=-\epsilon^{ijk} v_j B_k=-\frac{\epsilon^{ijk}\epsilon^{jlm} E_l B_m B_k}{\vec{B}^2}=E_i\;,\label{EvB}
\ee
so $\vec{E}+\vec{v}\times \vec{B}=0$, which means that if the electromagnetic field 
is Bateman's knot and the velocity field is the 
fluid knot derived from it, the source on the right-hand side of the basic equations (\ref{fluideqs}) is zero, 
leaving us with the usual Euler equation, that was already satisfied due to our map. 
That means that now we have a solution of the combined system, with nontrivial $\vec{v},\vec{E},\vec{B}$!

\subsection{Hopfion solution to Abanov-Wiegmann equations for the helicities}

As was shown above, the fluid + electromagnetic Hopfion solves the  continuity and Euler equations (\ref{fluideqs}) 
for the case of a divergent-less pressure $\nabla P=0$. Since the Abanov-Wiegmann equations are derived from 
them, it follows that the same fluid + electromagnetic Hopfion solves them as well.

But the condition $\nabla P=0$ is too restrictive, so we ask: can we relax it, if we consider only the 
Abanov-Wiegmann equations for the helicity densities? The answer turns out to be yes, as we now show.



By construction from the map (\ref{velmap}) we obtain easily that 
\be
\vec{B}\cdot\vec{v}=0\;,
\ee

Furthermore, since the EM field are null, we have (\ref{EvB})
which means that 
the Lorentz force (source of the 
fluid equations) vanishes, 
\be
\vec E+ \vec v\times \vec B =0.
\ee


Then the second  equation in the Abanov-Wiegmann equations (\ref{helicityeqs}) takes the form 
\be
\nabla\left[ \cdot\vec B\left (\mu-\frac12 mv^2  \right )\right ] =0.
\ee

Inserting the map for the velocity, we find that the second term vanishes and thus the equation holds for fluids for which
(note that $\vec{v}^2=1$ for the fluid solution constructed as above, so $\vec{\nabla} v^2=0$, and also $\vec{\nabla}\cdot\vec{B}=0$ 
from Maxwell's equations satisfied by the electromagnetic knot)
\be
\frac{1}{\rho}\vec B\cdot \vec\nabla p= 0.
\ee

Note that this condition is valid even for a non-barotropic fluid.

On the other hand from the first equation in (\ref{helicityeqs}), we get 
\be
\d_t(m^2\vec{v}\cdot\vec{\omega})+\vec{\nabla}\cdot\left[\vec{v}(m^2\vec{v}\cdot\vec{\omega})
+m\vec{\omega}\left(\mu-\frac{m\vec{v}^2}{2}\right)\right ].
\ee

The third term includes  two terms, one of which is proportional to 
$\vec \nabla\cdot\vec w=0$, and a term $\vec \nabla(v^2)=0$, so the equation takes the form
\be
\d_t(\rho_w)+\vec{\nabla}\cdot\left[\vec{v}( \rho_w)\right]
+m\vec{\omega}\cdot\vec\nabla\mu=0\;, \label{rhoomega}
\ee
where 
\be
\rho_w\equiv m^2\vec{v}\cdot\vec{\omega}\equiv h_0^f
\ee
is the fluid helicity density.

We now calculate explicitly the vorticity $\omega$ and $h_0^f$, and check explicitly the above equation. 
We find, as for the plain fluid Hopfion,
\bea
w_x &=& \frac{2 \left(t^2 y+2 t (x-y z)+y \left(x^2+z^2+3\right)-2 x z+y^3\right)}{\left((t-z)^2+x^2+y^2+1\right)^2}\CR
w_y &=& -\frac{2 \left(x \left((t-z)^2+y^2+3\right)+2 y (z-t)+x^3\right)}{\left((t-z)^2+x^2+y^2+1\right)^2}\CR
w_z &=& -\frac{4 \left((t-z)^2+1\right)}{\left((t-z)^2+x^2+y^2+1\right)^2}
\eea
and, for the velocity solution with minus sign in $v_z$ in (\ref{rhov}), we find
\be
\rho_w= \frac{4}{(t-z)^2+x^2+y^2+1}.
\ee

Inserting this expression into (\ref{rhoomega}), we find that

\be
\d_t(\rho_w)+\vec{\nabla}\cdot\left[\vec{v}( \rho_w)\right]
 =0.
\ee

Thus the first equation in (\ref{helicityeqs}) is fulfilled, provided  the pressure obeys
\be
\vec{\omega}\cdot\frac{\vec\nabla p}{\rho} = 0.
\ee

Again, we understand this to be true even in the non-barotropic case, since the terms with $\mu=\int dp/\rho$
have already cancelled.

Thus we conclude that indeed the null fluid solution combined with the null electromanetic configuration form a solution 
of the Abanov-Wiegman helicity equations, provided that the gradient of the pressure is perpendicular  to both $\vec B$ and $\vec w$, 
which means a potentially more general solution (with pressure) than the one for the (\ref{fluideqs}) equations.

We can ask: can the Hopfion itself be extended in the above way? The answer is that it cannot be extended 
with a nonconstant barotropic pressure, $p=p(\rho)$. For a null fluid obtained from electromagnetism
(so $\vec{E}\cdot \vec{B}=\vec{E}^2-\vec{B}^2$ and $\vec{\nabla}\cdot \vec{E}=\vec{\nabla}\cdot \vec{B}
=0$), we obtain 
\be
\vec{\omega}\cdot\frac{\vec{\nabla} p}{\rho}=\frac{dp}{\rho d\rho}\vec{\omega}\cdot \vec{\nabla \rho}
=\frac{dp}{\rho d\rho}\frac{1}{\vec{E}^2}\left((\vec{E}\cdot\vec{\nabla})\vec{B}-(\vec{B}\cdot \vec{\nabla})
\vec{E}\right)\cdot\vec{\nabla}\vec{E}^2\;,
\ee
which is nonzero in general. Moreover, even for the Hopfion solution, we obtain, at $t=0$, 
\be
\vec{\omega}\cdot\vec{\nabla}\rho=\frac{-16z}{(x^2+y^2+z^2+1)^6}\left[-3+\frac{4}{(x^2+y^2+z^2+1)^2
}\right]\neq 0.
\ee

So we can try for specific fluid+electromagnetic solutions other than the Hopfion, or otherwise consider 
a non-barotropic extension, i.e., consider $\vec{\nabla}p$ independent of $\vec{\nabla}\rho$.

\subsection{General knotted solution with nontrivial fluid-electromagnetic interaction}

In the solutions found until now, the Lorentz force, giving the interaction between the fluid and electromagnetism, was zero, 
so the solutions were not so satisfactory. We can ask: are there truly interacting solutions, such that the Lorentz force is nonzero, 
and there is a chance for the helicity to change between the fluid and electromagnetism?

The answer turns out to be yes, with a caveat. In the equations (\ref{fluideqs}), the Maxwell fields are considered as external 
fields, so strictly speaking they don't need to satisfy Maxwell's equations. More precisely, they need to satisfy the Bianchi 
identity, $\d_{[\mu }F_{\nu\rho]}=0$, which is just a statement that the electromagnetic fields are generated by a gauge field $A_\mu$, 
but they do not need to satisfy the equations of motion, which can be thought to be in the presence of some unknown (not 
considered in the fluid equations) source, 
\be
\d^\mu F_{\mu\nu}=\tilde j_\nu\neq 0.
\ee

Under this caveat, we can construct a solution as follows. 

First, calculate a velocity field from the electromagnetic knot solution in (\ref{solF}), using our electromagnetism to fluid map, 
\bea
v^i&=&\frac{\epsilon^{ijk} E^j B^k}{\frac{1}{2}(\vec{E}^2+\vec{B}^2)}
=\frac{2 F^{0j}F^{ij}}{(F^{0k})^2+(\frac{1}{2}\epsilon^{klm}F_{lm})^2}\cr
&=&\frac{2(\d^0\bar \phi\d^j\phi-\d^0\phi\d^j\bar\phi)(\d^i\bar\phi\d^j\phi-\d^i\phi\d^j\bar\phi)}{(\d^0\bar\phi\d^k\phi-\d^0\phi\d^k\bar\phi)^2
+(\epsilon^{klm}\d^l\bar\phi\d^m\phi)^2}.
\eea

This satisfies 
\bea
u^\mu F_{\mu\nu}&=&0:\cr
v^iF_{i0}&=&0\Rightarrow \vec{v}\cdot\vec{E}=0\cr
F_{0i}+v^jF_{ji}&=&0\Rightarrow -(\vec{E}+\vec{v}\times \vec{B})^i=0\;,\label{usol}
\eea
where $u^\mu =(1,v^i)$ and $F_{\mu\nu}$ is the solution in (\ref{solF}). 

But, if in (\ref{solF}) we replace on the left-hand side $F_{\mu}$ by 
\be
\Omega_{\mu\nu}=\d_\mu \pi_\nu-\d_\nu \pi_\mu=F_{\mu\nu}+{\cal V}_{\mu\nu}\;,
\ee
where
\be
\mathcal{V}_{0i}= m(\pa_0 v_i -v_j\pa_i v^j)\qquad \mathcal{V}_{ij}=m(\pa_i v_j-\pa_j v_i)\;,\label{vmunu}
\ee
meaning that we now have 
\bea
\Omega&=&\frac{1}{2}\Omega_{\mu\nu}dx^\mu\wedge dx^\nu
=\frac{1}{4\pi i}\frac{\d_\mu \bar\phi\d_\nu \phi-\d_\nu\bar\phi\d_\mu \phi}{(1+|\phi|^2)^2}
dx^\mu\wedge dx^\nu\cr
\tilde \Omega&=&\frac{1}{2}\tilde \Omega_{\mu\nu}dx^\mu\wedge dx^\nu
=\frac{1}{4\pi i}\frac{\d_\mu \bar\theta\d_\nu \theta-\d_\nu\bar\theta\d_\mu 
\theta}{(1+|\theta|^2)^2}dx^\mu\wedge dx^\nu\;,\label{Omegasol}
\eea
it follows that we now satisfy (remember that $j^\mu=(\rho,\rho v^i)=\rho u^\mu$)
\be
u^\mu \Omega_{\mu\nu}=0\Rightarrow j^\mu \Omega_{\mu\nu}=0\;,
\ee
i.e., the fluid Euler equations. Since the velocity field is defined by (\ref{usol}) and $\Omega_{\mu\nu}$ by (\ref{Omegasol}), 
we can subtract the velocity field ${\cal V}_{\mu\nu}$ from $\Omega_{\mu\nu}$, which will define the electromagnetic field 
$F_{\mu\nu}$. 

This solution will not satisfy  the vacuum equation of motion $\d^\mu F_{\mu\nu}=0$. 
But that is fine, since $\vec{E}$ and $\vec{B}$ were considered as external fields, not as Maxwell fields, 
so we can {\em define} the right-hand side as the current that generates them, so $\d^\mu F_{\mu\nu}\equiv \tilde j^\nu$. 

All we need to satisfy are the Bianchi identities for $F_{\mu\nu}$, and those are satisfied, since $\Omega_{\mu\nu}$ 
satisfies them by construction, $\d_{[\mu}\Omega_{\nu\rho]}=0$, and so does ${\cal V}_{\mu\nu}$, since 
\bea
{\cal V}_{ij}&=&m(\d_i v_j-\d_j v_i)\cr
{\cal V}_{0i}&=& m\d_0 v_i -\d_i \Phi=m\left(\d_0 v_i+\d_i \frac{\vec{v}^2}{2}\right)+\d_i \mu\;,
\eea
so obviously satisfies $\d_{[\mu}{\cal V}_{\nu\rho]}=0$, and therefore their difference, $F_{\mu\nu}$ also does:
\be
\d_{[\mu}F_{\nu\rho]}=0.
\ee


\section{ Euler-dual-Maxwell system of magnetically charged fluid}

In this section, we consider the (Poincar\'{e}) dualization of the previous construction, 
and then speculate about possible physical application. Before that, however,  we review the  
definition and properties of the dual electromagnetic theory.

\subsection{ Dualizing Maxwell equations}

In the absence of sources, the electromagnetic theory can be dualized. In a covariant formulation, the dual of 
\be
F_{\mu\nu}= \pa_\mu A_\nu - \pa_\nu A_\mu 
\ee
is 
\be
G_{\mu\nu}\equiv \frac{1}{2}\epsilon_{\mu\nu\rho\sigma}F^{\rho\sigma} = \pa_\mu C_\nu - \pa_\nu C_\mu .
\ee

The duality of the electric and magnetic fields is manifested by defining a field $\vec{C}$ in a similar way to $\vec{A}$,
\be
\vec{E}=\vec{\nabla}\times \vec{C};\;\;\; \vec{B}=\vec{\nabla}\times \vec{A}\;.
\ee

In the Bateman formulation \cite{Bateman:1915}, as used before, one defines the complex four-vector
\be
H_\mu \equiv \frac12( \alpha\pa_\mu \beta- \beta \pa_\mu \alpha)\;, \qquad C_\mu = {\rm Re}( H_\mu)\;, \qquad A_\mu= {\rm Im(} H_\mu).
\ee
 

\subsection{Conserved ``helicities"}

In electromagnetism, we can introduce helicities, that can be conserved under conditions to be defined shortly. 
We already introduced the magnetic helicity, that was part of the conserved total, fluid-electromagnetic, helicity ${\cal H}_{\rm tot}$, 
\be
{\cal H}_{mm}=\int d^3x \vec{A}\cdot \vec{B}=\int d^3x\epsilon^{ijk}A_i\d_j A_k.
\ee 

As we see, the helicities are spatial integrals of Chern-Simons forms, ${\cal H}_{mm}$ being the integral of the 
Chern-Simons form of $\vec{A}$. The electric-magnetic dual of the above is the 
electric helicity, defined as an integral of a Chern-Simons form of $\vec{C}$,
\be
{\cal H}_{ee}=\int d^3x \vec{C}\cdot \vec{E}=\int d^3x \vec{C}\cdot \vec{\nabla}\times \vec{C}=\int d^3x \epsilon^{ijk}C_i\d_j C_k\;.\label{Hee}
\ee

We can also define the integrals of BF forms, the cross helicities, the eletromagnetic one, 
\be
{\cal H}_{em}=\int d^3x \vec{C}\cdot\vec{B}=\int d^3x \epsilon^{ijk}C_i\d_j A_k\;,
\ee
and its electromagnetic dual, the magnetoelectric one,
\be
{\cal H}_{me}=\int d^3x \vec{A}\cdot\vec{E}=\int d^3x \epsilon^{ijk}A_i\d_k C_k.\;,
\ee
though of course for fields that vanish at infinity they are equal by partial integration.

The helicities are interesting, since although they are defined in terms of $\vec{A}$ and $\vec{C}$, 
they are gauge invariant under the 3-dimensional gauge transformations generated by $\a(\vec{x})$, 
since they are Chern-Simons and BF form. Of course, that is true only if 
the transformation of the Abelian fields $\vec{A}(\vec{x},t)$ is not a large gauge transformation, 
so that there are no global issues. Under large
gauge transformations, as is the case for any CS or BF integrals, they can change by an integer times $2\pi$. 

In the cases relevant for us, with time dependence, so  $\vec{A}(\vec{x},t)$, like we are considering in this paper, conservation 
of these helicities in time is not guaranteed, and neither is an integer value for them (though, as we saw, for the superfluid case, 
the total helicity ${\cal H}_{\rm tot}$, containing ${\cal H}_{mm}$, is quantized)

Using the Maxwell's equations and partial integrations, the time evolution of the electromagnetic helicities is found to be
\bea
\d_t {\cal H}_{mm}&=&\int d^3x (\d_t \vec{A}\cdot \vec{B}+\vec{A}\cdot \d_t\vec{B})
=-\int d^3x(\vec{E}\cdot\vec{B}+\vec{A}\cdot(\vec{\nabla}\times\vec{E}))\cr
&=&-2\int d^3x \vec{E}\cdot\vec{B}\cr
\d_t {\cal H}_{ee}&=&\int d^3x (\d_t \vec{C}\cdot \vec{E}+\vec{C}\cdot \d_t\vec{E})
=-\int d^3x(\vec{B}\cdot \vec{E}+\vec{C}\cdot (\vec{\nabla}\times \vec{B}))\cr
&=&-2\int d^3x \vec{E}\cdot \vec{B}\cr
\d_t {\cal H}_{me}&=& \int d^3x(\d_t \vec{A}\cdot \vec{E}+\vec{A}\cdot\d_t\vec{E})
=\int d^3x(-\vec{E}\cdot\vec{E}+\vec{A}\cdot(\vec{\nabla}\times\vec{B}))\cr
&=&-\int d^3x(\vec{E}^2-\vec{B}^2)\;,\cr
\d_t {\cal H}_{em}&=& \int d^3x(\d_t \vec{C}\cdot \vec{B}+\vec{C}\cdot \d_t \vec{B})
=\int d^3x(-\vec{B}\cdot \vec{B}+\vec{C}\cdot (\vec{\nabla}\times \vec{E}))\cr
&=&\int d^3x (\vec{E^2}-\vec{B}^2)\;.
\eea

 Thus  the helicities  $H_{mm}$ and $H_{ee}$  are  conserved for configurations  for which  
 $\vec{E}\cdot \vec{B}=0$, and  $H_{em}$ and $H_{me}$ are conserved provided that   $\vec{E^2}-\vec{B}^2=0$. 
For the null configurarions defined in (\ref{null}) 
these two conditions are obeyed, so 
for them all the four helicities are conserved.


\subsection{Magnetically charged fluid}

In the case of the Euler fluid coupled to electromagnetism via electric charges, i.e., a fluid made up of electrically charged
particles, we saw that only the total helicity, fluid plus ${\cal H}_{mm}$, is conserved. 


But we can also consider a fluid of magnetically charged particles, where the source in the Euler equation is the 
magnetically charged Lorentz force, so it obeys the equations
\bea
&&\frac{d}{dt}\rho+\vec{\nabla}\cdot \rho\vec{v}=0\cr
&&(\d_t+\vec{v}\cdot\vec{\nabla})m\vec{v}+\vec{\nabla}\mu=g\vec{B}-g\vec{v}\times\vec{E}\;.\label{fluideqs2}
\eea

Formally, the derivation of \cite{Abanov:2021hio}, reviewed in section 2, can be repeated for a ``magnetically charged'' fluid, 
by using the electric-magnetic dual conjugate momentum,
\be
\vec\pi_c = m\vec v + \vec C.
\ee

Now the total  helicity of the fluid plus the electric helicity $ 
{\cal H}_{ee}$ is defined as 
\be
{\cal H}^m_{\rm tot} =\frac{1}{h^2}\int d^3 x \vec \pi_c \cdot \nabla\times \pi_c
\ee
so that 
\be
{\cal H}^m_{\rm tot}=\frac{1}{h^2}\int d^3x \left[m\vec{v}\cdot\vec{\omega}+\vec{C}\cdot\vec{E}+2m\vec{v}\cdot\vec{E}\right]
={\cal H}_f+{\cal H}_{ee}+2{\cal H}_{fe}.
\ee

Following the same steps of the derivation of the equations for electrically charged fluid, we find for the magnetically charged one 
that the new total helicity ${\cal H}^m_{\rm tot}$ is conserved, and moreover
we find  the analog of the Abanov-Wiegmann equations,
\bea
&&\d_t(m^2\vec{v}\cdot\vec{\omega})+\vec{\nabla}\cdot\left[\vec{v}(m^2\vec{v}\cdot\vec{\omega})
+m\vec{\omega}\left(\mu-\frac{m\vec{v}^2}{2}\right)+m\vec{v}\times(\vec{B}-\vec{v}\times \vec{E})\right]\cr
&&-2m\vec{\omega}(\vec{B}
-\vec{v}\times \vec{E})=0\cr
&&\d_t(m\vec{v}\cdot\vec{E})+\vec{\nabla}\cdot \left[\vec{v}(m\vec{v}\cdot\vec{E})+\vec{E}\left(\mu-\frac{m\vec{v}^2}{2}\right)
-m\vec{v}\times (\vec{B}-\vec{v}\times \vec{E})\right]\cr
&&+m\vec{\omega}(\vec{B}-\vec{v}\times \vec{E})=\vec{B}\cdot\vec{E}\;.\cr
&&\label{helicityeqsDual}
\eea

\subsection{Possible physical applications}

It may seem that it is a bit abstract to talk about a fluid of magnetically charged particles, 
as in the previous subsection, when magnetic monopoles haven't
even been observed yet. One possibility would be that we will find somewhere in the Universe some place where magnetic monopoles
abound and form a fluid, and that is certainly one application of the previous formalism.

However, really, what we want are {\em effective} particles. So, for instance, if we consider 
a type II superconductor close to the phase transition, it will be composed of many parallel magnetic flux tubes, almost overlapping. 

So from the point of view of the 2+1 dimensional reduced theory, the flux tubes look like particles, with some radius, interacting, which is 
kind of a description of a fluid anyway. So we can consider a 2+1 dimensional fluid of such tubes in the superconductor, interacting with 
external {\em electric } fields, to be governed by the above description.

Hence the above eletric/magnetic case also be dimensionally reduced to 2+1 dimensions 
in exactly the same way as we do for the
${\cal H}_{mm}$ helicity, and we can use it for the "effective fluid" of flux tubes in a type II superconductor. 

However, there is a trick when reducing this dual, magnetically charged fluid to 2+1 dimensions. 
In 2+1 dimensions, $B$ is a scalar, understood to be transverse to the 2-dimensional plane, so the 
magnetic Lorentz force becomes a scalar as well,
\be
g(B-\epsilon^{ab}v_a E_b)\;,
\ee
and, since this is understood as the force in the perpendicular direction, where for a consistent 
reduction the left-hand side in (\ref{fluideqs2}) must be zero, which means that the right-hand side 
must be zero as well, so the above Lorentz force must vanish. 

Then, for this consistent reduction, we have that the (quantized) magnetic flux is understood as 
a 2+1 dimensional helicity involving the electric field, 
\be
\Phi_m=\frac{e}{h}\int d^2x B=\frac{1}{g}\int d^2 x \epsilon^{ab}v_a E_b\equiv {\cal H}_e\;,
\ee
where we also used Dirac quantization, $eg=h$.

Thus we reinterpret the 2+1 dimensional case of 
type II superconductor as a dual magnetically charged fluid, with this electric helicity ${\cal H}_e$.

\section{Non-linear generalization of the Abanov Wiegmann system}


In \cite{Alves:2017ggb,Alves:2017zjt} it was proven  that not only are the knotted solutions 
like the Hopfion solutions of Maxwell's theory, they are also solutions of 
{\em any} nonlinear theories that reduce to electromagnetism at small fields, theories like the Born-Infeld theory, 
or theories obtained by integrating 
out {\em any} light fields interacting with electromagnetism, like the Euler-Heisenberg Lagrangian. This was also  discussed  
in \cite{Goulart:2016orx}.  

Based on this insight and the fact that, as we have seen,  the electromagnetic Hopfion and the corresponding velocity in 
(\ref{rhov}) form a solution of the Abanov-Weigmann equations, we would like now to generalized these equations of 
motion to those that one gets upon replacing Maxwell's theory  with one of its non-linear generalizations. 

We follow here the steps taken in \cite{Alves:2017ggb,Alves:2017zjt} and we start with the generalization of the Maxwell 
electromagnetic theory to the 
  the formalism of Born and Infeld \cite{Born:1934gh}   (see also \cite{Nastase:2015ixa}   for 
a generalization of this analysis). Defining the quantities
\bea
F&\equiv& \frac{F_{\mu\nu} F^{\mu\nu}}{2b^2}=\frac{1}{b^2}(\vec{B}^2-\vec{E}^2);\cr
G&\equiv&
\frac{1}{8b^2}\epsilon^{\mu\nu\rho\sigma}F_{\mu\nu}F_{\rho\sigma}= \frac{1}{b^2}\vec{E}\cdot \vec{B}\;,
\eea
where $b$ is a dimensionful constant of mass dimension 2, the BI Lagrangian is 
\be
{\cal L}=-b^2[\sqrt{1+F-G^2}-1].
\ee

We define the conjugate quantities analogous to the ones of electromagnetism in a medium,
\be
\vec{H}\equiv -\frac{\d {\cal L}}{\d \vec{B}},\;\;\;\;
\vec{D}\equiv +\frac{\d {\cal L}}{\d \vec{E}}\;.
\ee

The Maxwell equations in terms of them take the same form as for electromagnetism in a medium,
\bea
\vec{\nabla}\times \vec{E}+\d_0 \vec{B}=0, && \vec{\nabla}\cdot \vec{B}=0\cr
\vec{\nabla}\times\vec{H}-\d_0\vec{D}=0, && \vec{\nabla}\cdot \vec{D}=0.
\eea

In the BI theory case, we have 
\be
\vec{H}=\frac{\vec{B}-G\vec{E}}{\sqrt{1-F+G^2}};\;\;\; 
\vec{D}=\frac{\vec{E}+G\vec{B}}{\sqrt{1-F+G^2}}\;.
\ee

We see explicitly that for $F=G=0$, like we have for the Hopfion and for the null knotted solutions, we obtain $\vec{H}=\vec{B}$ and 
$\vec{D}=\vec{E}$, and therefore the Maxwell equations reduce to the ones in vacuum. 

From the relations $\vec{H}(\vec{E},\vec{B}), \vec{D}(\vec{E},\vec{B})$, we find that the same is true for any 
Lagrangian that is only a function of $F$ and $G$, and which contains electromagnetism as a small fields limit, 
i.e., a Lagrangian that can be written as an expansion
containing the Maxwell term plus higher order terms,
\be
{\cal L}=b^2\left[-\frac{F}{2}+\sum_{n\geq 2}\sum_{m\geq 0}c_{n,m}F^nG^m\right].\label{expan}
\ee

Indeed, in this case we obtain 
\be
\vec{H}=\vec{B}+{\cal O}(F,G);\;\;\;
\vec{D}=\vec{E}+{\cal O}(F,G)\;,
\ee
therefore if $F=G=0$ we obtain the usual Maxwell's equations in terms of $\vec{E}$ and $\vec{B}$.

However, introducing the replacement
\be
\vec E\rightarrow \vec D\;, \qquad \vec B\rightarrow \vec H\;,
\ee
in the Maxwell's equations,
given that the particle coupling to electromagnetism is still $q\int dx^\mu A_\mu$, leading to the same 
$qF_{\mu \nu}u^\nu$ Lorentz force, so in terms of the same $\vec{E},\vec{B}$, 
the DBI- Euler equations of motion are actually the same as in (\ref{fluideqs}),
and from them we can find the same Abanov-Wiegmann equations (\ref{helicityeqs}).


\section{Two-dimensional formalism and solutions}

Next, we consider the 2+1 dimensional version of the fluid equations, and of the 4 dimensional fluid formalism, and seek equations 
to them.

\subsection{2+1 dimensional generalization of the 4 dimensional formalism and equations}

The 2+1 dimensional version of the continuity and Euler equations is 
\bea
&&\d_t \rho+\d^a(\rho v_a)=0\cr
&&(\d_t+v^c\d_c)m v^a+\d^a \mu=eE^a+\frac{e}{c}\epsilon^{ab}v_b B.\label{2dEuler}
\eea

Note that in 2 spatial dimensions we have
\be
E^a=F^{0a}=-(\d_0 A_a-\d_a A_0)\;,\;\;
B=\epsilon^{ab}\d_a A_b.
\ee

That means, since $\omega=\epsilon^{ab}\d_a v_b, B=\epsilon^{ab}\d_a A_b$, that the only possible definition of a 
(reduced) 2+1 dimensional helicity is
\be
H_{\rm tot, red}=\frac{1}{h}\int d^2x \epsilon^{ab}\d_a \pi_b=\int d^2x (m\omega+eB).
\ee

Note that the normalization was chosen such that the purely magnetic helicity, equal to the magnetic flux through the plane, has integer 
values. Indeed, we know that the fluxon is $\Phi_0=h/e$, so $\frac{e}{h}\int d^2 x B=k\in \mathbb{Z}$.

Then, construct
\bea
\pi_a&=& mv_a+eA_a\cr
\pi_0&=&\Phi+eA_0\cr
-\Phi&=&\mu+\frac{mv^2}{2}.
\eea

The Euler equation is rewritten as 
\be
\rho(\d_t \pi^a-\d^a\pi_0)-\rho \epsilon^{ab}v_b(m\omega+eB)=0\;,
\ee
as we can easily check (use $\epsilon^{ab}\epsilon^{cd}=\delta^a_c\delta^b_d-\delta^a_d\delta^b_c$),
where 
\be
\vec{\nabla}\times\vec{\pi}\rightarrow \epsilon^{ab}\d_a\pi_b={\cal H}_{\rm tot}^d=m\omega+eB.
\ee

This rewriting of the Euler equation can be further written in a compact form, first defining 
\be
\Omega_{\mu\nu}=\d_\mu \pi_\nu-\d_\nu \pi_\mu\;,
\ee
and then $j^\mu=(\rho,\rho v^a)$, as before,
\be
j^\mu \Omega_{\mu\nu}=0.
\ee

The continuity equation is $\d_\mu j^\mu=0$, also as before. 

We define the helicity density 2-form and its 1-form dual, 
\be
h=*d\pi=*\Omega.
\ee

Then the total helicity density is its 0 component,
\be
{\cal H}_{\rm tot}^d=h_0=\frac{1}{2}\epsilon^{ab}\Omega_{ab}=\epsilon^{ab}\d_a\pi_b\;,
\ee
and the helicity flux are the $a$ components,
\be
\vec{h}: h^a=\epsilon^{ab}\Omega_{0a}=\epsilon^{ab}(\d_0\pi_a-\d_a\pi_0).
\ee

The conservation of helicity is once again trivial, 
\be
d*h=dd\pi=0\;,
\ee
in components
\be
\dot h_0+\d^a h_a=0\Rightarrow \d_t\int d^2x h_0=0\;,
\ee
with appropriate boundary conditions at infinity. 

Now define the time dependence of the electromagnetic helicities. 
As we said, in 2+1 dimensions, we can define the magnetic helicity which is just magnetic flux, 
\be
H_m=\frac{e}{h}\int d^2x B\;,
\ee
but, unlike in 3+1 dimensions, there are no other possibilities, since the electric field $\vec{E}$, unlike $B$, is a vector, 
so cannot be used to contract with $B$. So we cannot even define a dual helicity $H_e$, since in 
2+1 dimensions, the dual to the vector $A_\mu$ is a scalar, call it $C$ (similar to the $\vec{C}$ in 4 dimensions). 

Then, the time derivative of $H_m$ (which in 4 dimensions 
is proportional to $\int d^3x \vec{E}\cdot\vec{B}$) is now, 
via the 2+1 dimensional Maxwell's equations, 
\be
\frac{d}{dt}H_m\propto\frac{d}{dt}\int_S d^2x B d S=-\int_S d^2x\vec{\nabla}\times \vec{E} dS=-\oint_{C=\d S}\vec{E}\cdot d\vec{l}\;,
\ee
and in 2+1 dimensions 
\be
\vec{\nabla}\times \vec{E}=\epsilon^{ab}\d_a E_b.
\ee

\subsection{2+1 dimensional reduction of solutions}

In \cite{Alves:2017ggb,Alves:2017zjt}, it was also shown how to dimensionally reduce a fluid solution (of the Euler equation) 
to 2+1 dimensions. 

First, one notices that the velocity for the fluid knot (\ref{rhov}), obtained from the map from the electromagnetic knot, can be  
rewritten in general (for $t-z\neq 0$) as the vortex-like solution
\be
v^a=\epsilon^{ab}\partial_b \psi+(t-z) \partial^a\psi,\ \ \psi=\log(1+x^2+y^2+(t-z)^2).\label{nullvortex}
\ee

For it, the (reduced) 2+1 dimensional fluid helicity, the integral of the vorticity, is constant,
\be
H_{\rm f, red}=\int d^2x \omega=4\pi\;,
\ee
though it should be written in units of $\frac{h}{4\pi m}$, for consistency with the magnetic flux. 

This fluid knot ("Hopfion") solution satisfies the continuity equation $\d_t \rho+\d_i(\rho v_i)=0$, $\d_+ v^a=0$ ($x^\pm =t\pm z$) and 
\be
\d_- v^a+\b^b \d_b v^a=0\;,\;\; \b^a=\frac{v^a}{1-v_z}.\label{Eulerred}
\ee

Here $\d_-$ now stands for $\d_t$, and taking $\epsilon^{ab}\d_b $ on the above equation, we obtain $\d_-\omega+\d_a(\epsilon^{ab}
\b^c\d_c v_b)=0$, which when integrated gives $\d_-\int d^2x \omega+\oint_{S_\infty}(...)=0$, or $\d_t H_{\rm f, red}=0$. That is, 
indeed the 2+1 dimensional fluid helicity is conserved in (reduced) time, $x^-$.

For a consistent reduction of the equations satisfied by the fluid knot to 2+1 dimensional Euler fluid equations, we have two options:

1. The first one is to multiply the equation  in (\ref{Eulerred}) by $(1-v_z)$, and note that 
we can define $\d^a p\equiv (1-v_z)\d_- v^a$ and consider constant density $\rho=1$, 
in which case we get the usual {\em static} (in $x^-$ time) Euler equation with pressure in 2+1d, 
\be
v^b \d_b v^a+\d^a p=0\;,
\ee
leading to the solution
\be
v_x=\frac{2y}{1+x^2+y^2}\;,\;\;
v_y=\frac{-2x}{1+x^2+y^2}\;,\;\;
p=p_\infty-\frac{2}{1+x^2+y^2}.\label{presssol}
\ee

2. The other option is to define new velocities, 
\be
\b^a=\frac{v^a}{1-v^z}\;,
\ee
in terms of which we have the continuity and Euler equations with constant pressure in 2+1 dimensions, in terms of $x^-$ time,
\be
\d_- \rho+\d_a(\rho \b^a)=0\;,\;\;
\d_-\b^a+\b^b\d_b\b^a=0\;,
\ee
integrated to ($\rho$ the same as in (\ref{rhov}) and)
\be
\beta^a=\epsilon^{ab}\partial_b \tilde{\psi}+(t-z)\partial^a\tilde{\psi},\ \ \tilde{\psi}=\log(x^2+y^2-1-(t-z)^2).\label{betasol}
\ee

Note the signs different inside the log in $\tilde \psi$ with respect to $\psi$ in (\ref{nullvortex}).

To find knot (or rather, vortex) solutions of the 2+1 dimensional equations (\ref{2dEuler}) solutions, 
as in 3+1 dimensions, it would suffice if we would have zero Lorentz force,
\be
E^a+\epsilon^{ab}v_b B=0.
\ee

This would be obtained from the 2+1 dimensional version of the fluid map, namely if we would have $E^aE^a=B^2$ 
(note that since $B$ basically means $B_z$, by definition we have $\vec{E}\cdot \vec{B}=0$) and 
\be
v^a=\frac{\epsilon^{ab}E_b}{\sqrt{E^cE_c}}.
\ee

Of course, while at case 1 we would have velocities $v^a$, at case 2 we would have velocities $\b^a$, and in both cases 
time is $x^-=t-z$.

\section{Solutions via conformal transformations}

In \cite{Hoyos:2015bxa} it was shown how to construct knotted solutions form un-knotted  solutions 
like constant and plane wave electric and magnetic fields  via 
complex special  conformal transformations.
The proof of this statement and  several applications  to construct knotted solutions was done using
 the Bateman variables\cite{Hoyos:2015bxa}.
Whereas the special conformal  transformation of the electric and magnetic field  are complicated,
 in terms of $\alpha$ and $\beta$ one has to perform only the transformation of the coordinates. 

The prototype  example is the  derivation of  the basic Hopfion solution  (\ref{absol}) from the  
configuration of constant perpendicular electric and magnetic fields given by  
\be
\alpha=2i(t+z) \qquad \beta=2(x-iy)\;,
\ee
which corresponds to electromagnetic fields
\be
\vec E =(-4,0,0) \qquad \vec B =(0,4,0).
\ee

Under the special conformal transformation
\be
x^\mu\rightarrow \frac{x^\mu + b^\mu x_\nu x^\nu}{1+ 2 b^\mu x_\mu + 
b_\mu b^\mu x_\nu x^\nu}\;,\label{SCT}
\ee
with 
\be
b^\mu = i( 1,0,0,0)\;,
\ee
we obtain the  Hopfion solution  written in (\ref{absol}), namely with
\be
\a=\frac{A-iz}{A+it}\;,\;\;
\b=\frac{x-iy}{A+it}\;,\;\;
A=\frac{1}{2}(x^2+y^2+z^2-t^2+1).
\ee

Since, as was explained in section 3, we construct the velocity $\vec v$ from the electric and 
magnetic fields, then we can also     generate novel knotted fluid configurations by applying 
complex  special conformal transformations on $\alpha$ and $\beta$ that yields novel EM knot
 configuration from which we get  the fluid configurations.

We now demonstrate this method by deriving the $(p,q)$ knotted EM and  the corresponding fluid configurations.   
For this case we start with 
\be
\alpha=[2i(t+z)]^p \qquad \beta=[2(x-iy)]^q\;.
\ee

Next we apply the same special conformal transformation as the one given above in (\ref{SCT}), to find
\be
\alpha=\frac{\left(-t^2+x^2+y^2+(z+i)^2\right)^p}{\left(-t (t-2
   i)+x^2+y^2+z^2+1\right)^p}
\ee
\be
\beta= \frac{8 (x-i y)^q}{\left(-t (t-2 i)+x^2+y^2+z^2+1\right)^q}.
\ee

The corresponding fluid density  
 at $t=0$ has a denominator that is of the form \\$\left(x^2+y^2+z^2+
   1\right)^{2(p+q)}$.
 
A similar path can be applied on the the variables  $\alpha$ and $\beta$  that corresponds to an EM plane-wave,
 \be
\alpha= e^{i(z-t)}\ \ ; \qquad  \beta= x+iy\;,
\ee
which after the above complex special conformal transformation reads
\be
\alpha=\exp\left(-1 +\frac{i(t+z-i)}{(2A+i t)}\right)\ \ ; \qquad  \beta= \frac{x+iy}{(2A+i t)}.
\ee



In a similar manner to applying complex SCT on the basic configurations of the constant and 
plane wave electric and magnetic fields, one can also apply  conformal transformations  on the Hopfion itself,  
in particular time translations, space-translations, rotations, boosts and scale transformations with 
imaginary parameter. 


\section{Conclusions and discussion}

In this paper we have considered knotted solutions to the Euler fluid plus external electromagnetism equations, and to the 
equations for the helicities of the fluid and electromagnetism (AW equations). We have found that a map from electromagnetism to a null
fluid can be used to find knotted solutions, with helicities, to the coupled system, and to the Abanov-Wiegmann helicity equations. 
An electromagnetic dual case, for a magnetically charged fluid, was found to be similar. 
The case of nonlinear electromagnetism was treated similarly, with similar results. 
The 3+1 dimensional formalism for the fluid, used in the derivation of the conservation of the total helicity, was 
extended to 2+1 dimensions, and the solutions in 3+1 dimensions were dimensionally reduced, to find 2+1 dimensional 
solutions. Using conformal transformations, we were able to obtain the knotted solutions from unknotted ones.

There are many open questions related to this topic. Here we list several of them:
\begin{itemize}
\item
The systems discussed here did not incorporate the back-reaction of the fluid on the EM fields. The latter were taken to be external fields. It will be very interesting to find fluid and EM configurations that solve the coupled equations. This is of particular interest since so far all the EM knots have been solutions only to the free Maxwell's equations with not currents and charges.
\item
The fluids involved in the knotted solutions did not permit gradients of the pressure. It will be interesting to explore the possibity to find fluid knots in the presence of non-constant pressure. 
\item
A very interesting question is whethere one can also construct gauge and fluid knots associated with non-abelian gauge symmetry. This implies searching  for solutions of the YM equation  generalizing the AW equations to incorporate non abelian charges.
\item
EM knots  can easily be constructed using special conformal transformations with imaginary parameters. It is very natural in the context of Bateman formulation of the EM theory. It will be interesting to study these transformation in a anaologous formulation of fluid dynamics. 
\end{itemize}

\section*{Acknowledgements}
We would like to thank A.G. Abanov,   for useful comments and discussions and comments on 
the manuscript, and to Carlos Hoyos for several 
useful comments on the manuscript.

The work of HN is supported in part by  CNPq grant 301491/2019-4 and FAPESP grants 2019/21281-4 
and 2019/13231-7. HN would also like to thank the ICTP-SAIFR for their support through FAPESP grant 2016/01343-7.
The work of J.S was supported  by a center of excellence of the Israel Science Foundation (grant number
2289/18)
 
\appendix

\section{The (2,3) EM fluid knot solution}
Performing the complex special conformal transformations discussed in section 7 we derive 
the components of the electric and magnetic fields. At $t=0$ they are given by
\bea
E_x &=&   \frac{96 \left(-x^6+5 x^4 y^2-4 x^3 y z+x^2 \left(5 y^4+6 y^2
   \left(z^2-1\right)+z^4-6 z^2+1\right)\right)}{\left(x^2+y^2+z^2+1\right)^6}\cr
   &&+\frac{96\left(4 x y z \left(3 y^2+2 z^2-2\right)-y^2
   \left(2 \left(y^2-3\right)
   z^2+\left(y^2-1\right)^2+z^4\right)\right)}{\left(x^2+y^2+z^2+1\right)^6}\cr
E_y &=&  -\frac{192 \left(2 x^5 y-2 x^4 z+3 x^3 y \left(z^2-1\right)-2 x^2 z \left(-3
   y^2+z^2-1\right)\right)}{\left(x^2+y^2+z^2+1\right)^6} \cr
   &&-\frac{192\left(x y \left(-2 y^4-\left(y^2+6\right) z^2+y^2+z^4+1\right)+2
   y^2 z \left(z^2-1\right)\right)}{\left(x^2+y^2+z^2+1\right)^6}                                \CR
	E_z &=& -\frac{192 \left(x z^3 \left(x^2-3 y^2\right)-3 y z^2 \left(y^2-3 x^2\right)\right)}{\left(x^2+y^2+z^2+1\right)^6} \cr
	&&-\frac{192\left(x z
   \left(x^2-3 y^2\right) \left(x^2+y^2-3\right)-y \left(y^2-3 x^2\right)
   \left(x^2+y^2-1\right)\right)}{\left(x^2+y^2+z^2+1\right)^6} \CR
	B_x &=& \frac{192 \left(2 x^5 y+x^3 y \left(z^2-1\right)+2 x^2 z \left(3
   y^2+z^2-1\right)\right)}{\left(x^2+y^2+z^2+1\right)^6}\cr
   &&+\frac{192\left(-x y \left(2 y^4+3 y^2 \left(z^2-1\right)+z^4-6 z^2+1\right)-2
   y^2 z \left(y^2+z^2-1\right)\right)}{\left(x^2+y^2+z^2+1\right)^6}\CR
	B_y &=&  -\frac{96 \left(x^6+x^4 \left(-5 y^2+2 z^2-2\right)+12 x^3 y z+x^2 \left(-5 y^4-6
   \left(y^2+1\right) z^2+6 y^2+z^4+1\right)\right)}{\left(x^2+y^2+z^2+1\right)^6}\cr
   &&-\frac{96\left(-4 x y z \left(y^2-2 z^2+2\right)+y^2
   \left(y^4-z^4+6 z^2-1\right)\right)}{\left(x^2+y^2+z^2+1\right)^6}\CR
B_z &=&	-\frac{192 \left(x^5-3 x^4 y z+x^3 \left(-2 y^2+3 z^2-1\right)+x^2 y z \left(-2
   y^2-3 z^2+9\right)\right)}{\left(x^2+y^2+z^2+1\right)^6}\cr
   &&-\frac{192\left(-3 x y^2 \left(y^2+3 z^2-1\right)+y^3 z
   \left(y^2+z^2-3\right)\right)}{\left(x^2+y^2+z^2+1\right)^6}.\CR	
\eea

\bibliography{EulerMaxwellHopf}
\bibliographystyle{utphys}

\end{document}